\documentclass[pre,showpacs,twocolumn,superscriptaddress]{revtex4}

\usepackage{graphicx}
\usepackage{amssymb}
\usepackage{amsmath}
\usepackage{amsfonts}
\usepackage{mathrsfs}
\usepackage[usenames]{color}

\newcommand {\be} {\begin{equation}}   
\newcommand {\ee} {\end{equation}}
\newcommand {\bea} {\begin{eqnarray}}
\newcommand {\eea} {\end{eqnarray}}
\newcommand {\bes} {\begin{displaymath}}
\newcommand {\ees} {\end{displaymath}}
\newcommand {\beas} {\begin{eqnarray*}}
\newcommand {\eeas} {\end{eqnarray*}}

\begin{document}

\title{Breathing dynamics in heteropolymer DNA}

\author{Tobias Ambj\"ornsson}
\email{ambjorn@nordita.dk}
\affiliation{NORDITA (Nordic Institute for
Theoretical Physics), Blegdamsvej 17, DK-2100 Copenhagen \O, Denmark.}
\affiliation{Department of Chemistry,
Massachusetts Institute of Technology, 77 Massachusetts Avenue, Cambridge,
MA 02139, USA}
\author{Suman K. Banik}
\affiliation{Dept. of Physics, Virginia Polytechnic Institute and State
University, Blacksburg, VA 24061-0435, USA}
\author{Oleg Krichevsky}
\affiliation{Physics Department, Ben Gurion University, Be'er Sheva 84105,
Israel}
\author{Ralf Metzler}
\email{metz@uottawa.ca}
\affiliation{NORDITA (Nordic Institute for
Theoretical Physics), Blegdamsvej 17, DK-2100 Copenhagen \O, Denmark.}
\affiliation{Department of Physics, University
of Ottawa, 150 Louis Pasteur, Ottawa, Ontario  K1N 6N5, Canada}

\begin{abstract}
While the statistical mechanical description of DNA has a long tradition,
renewed interest in DNA melting from a physics perspective
is nourished by measurements of the fluctuation
\emph{dynamics\/} of local denaturation bubbles by single molecule
spectroscopy. The dynamical opening of DNA
bubbles (DNA breathing) is supposedly crucial for biological
functioning during, for instance, transcription initiation and DNA's
interaction with selectively single-stranded DNA binding proteins. Motivated
by this, we consider the bubble breathing dynamics in a heteropolymer
DNA based on a (2+1)-variable master equation and complementary
stochastic Gillespie simulations, providing the bubble size and
the position of the bubble along the sequence as a function of time.
We utilize new experimental data that independently obtain stacking and
hydrogen bonding contributions to DNA stability. We
calculate the spectrum of relaxation times and the
experimentally measurable autocorrelation function of a
fluorophore-quencher tagged base-pair, and demonstrate good agreement
with fluorescence correlation experiments. A significant
dependence of opening probability and waiting time between bubble
events on the local DNA sequence is revealed and quantified for a
promoter sequence of the T7 phage. The strong
dependence on sequence, temperature and salt concentration for the
breathing dynamics of DNA found here points at a good potential for
nanosensing applications by utilizing short fluorophore-quencher
dressed DNA constructs.
\end{abstract}


\maketitle 

\section{Introduction}


Textbook pictures of double-stranded DNA molecules may lead one to
believe that the Watson-Crick double-helix represents a static
geometry. As a matter of fact, even at room temperature DNA opens up
intermittent flexible single-stranded domains, so-called
\emph{DNA-bubbles}. Their size typically ranges from a few broken
base-pairs (bps), increasing to some 200 broken bps closer to the
melting temperature
$T_m$\cite{kornberg,poland,wartell,cantor_schimmel}. The stability of
DNA is characterized by the two free energies $\epsilon_{\rm hb}$ for
breaking the Watson-Crick hydrogen bonds between complementary AT and GC bps,
and the ten independent stacking free energies $\epsilon_{\rm st}$ for
disrupting the interactions between neighboring bps; at 100 mM NaCl
concentration and temperature 37 $^\circ$C it was found that
$\epsilon_{\rm hb}=1.0 k_BT$ for a single AT and $0.2 k_BT$ for a
single GC-bond (at $T=37 ^\circ$C, $k_BT=0.62$kcal/mol).
Under the same conditions the weakest (strongest) stacking
interaction was found to be the TA/AT (GC/CG) with free energies
$\epsilon_{\rm st}=-0.9 k_BT$ ($-4.1 k_BT$) \cite{FK1}.
In addition, the initiation
of a bubble in an unperturbed DNA molecule, creating two interfaces between
single-stranded bubble and vicinal double-helix at the zipper forks, is
associated with an activation factor $\sigma_0\simeq
10^{-3}\ldots 10^{-5}$ \cite{poland,wartell,blake,blossey} related to
the ring-factor $\xi$ used in \cite{FK,FK1} and below. That is, despite
the rather low free energy for breaking the bps, the high bubble nucleation
barrier
guarantees that below $T_m$ bubbles are rare and well separated, in particular,
under physiological conditions. However, once a bubble opens, since typical
free energies are of the order $k_BT$ localized denaturation zones can open
up, predominantly in AT-rich regions
\cite{wartell,poland,kornberg}. These DNA-bubbles fluctuate in size,
the \emph{DNA-breathing}. It has been demonstrated recently by
fluorescence correlation methods that DNA-breathing can be probed on
the single molecule level, revealing a multistate kinetics of stepwise
(un)zipping of bps with a bubble lifetime ranging up to a few
milliseconds \cite{altan}.

Theoretically, based on the statistical mechanical Poland-Scheraga model
\cite{poland} DNA-breathing has been described in homopolymer DNA in
terms of a continuous Fokker-Planck equation \cite{Hanke_Metzler}, and
through a stochastic Gillespie scheme \cite{suman}. A
discrete master equation approach was developed in \cite{tobias,tobiaslong},
including the coupled (un)binding dynamics of selectively
single-stranded DNA binding proteins. Continuous and discrete approaches
are compared and studied in \cite{bicout}.
Heteropolymer DNA-breathing was
considered in a reduced one-variable approach using a random energy
model \cite{hwa}.

Here, we develop a full $(2+1)$-variable approach to breathing in
heteropolymer DNA that allows us to study the sequence dependence of
the dynamics, through the initiation and the stochastic motion of the
two forks of an open DNA bubble.
Two approaches are used: the stochastic motion is
obtained by generating stochastic (Gillespie) time series from which
equilibrium distribution as well as autocorrelation functions are
obtained. We also use the corresponding master equation in order to
calculate the complementary ensemble-averages; excellent agreement is
found between time-averages and ensemble averages. Novelties in our
study include: (i) We study the full dynamics (2+1 variable problem)
of a heteropolymer region of {\em arbitrary} (not just random)
sequence; (ii) we compare our results to the fluorescence correlation
spectroscopy (FCS) experiments in
\cite{altan} using the directly measured DNA parameters in \cite{FK}
(see below); (iii) Recently, for the first time, the (two)
hydrogen bond energies, and (ten) stacking interactions characterizing
DNA stability within the Poland-Scheraga model were separately
determined \cite{FK,FK1}; these stability parameters are utilized in our
study. Among the consequences of these new results compared to previously
used parameters \cite{blake} are the more pronounced sequence
dependence and the fact that Watson-Crick and stacking interactions can
completely be separated as required when studying internal bubble
dynamics (a bubble involving $m$ broken Watson-Crick bonds and $m+1$
broken stacking interactions).

Based on this new approach, we study the question of transcription initiation
at the TATA motif of the biological sequence in the bacteriophage T7 promoter
sequence. Using the newly obtained stacking parameters from \cite{FK},
we demonstrate the delicate dependence of both the equilibrium opening
probability as well as the breathing dynamics on the sequence
dependence of the stacking. While in our model the opening times of bubbles
only marginally depend on their position along the sequence, the recurrence
frequence of bubble events is much more sensitive to the position. The
latter might therefore be a clue toward the understanding of transcription
initiation.

This paper is organized as follows: In Section \ref{sec:General model}
we describe the DNA bubble dynamics in terms of the relevant transfer
coefficients. In Section \ref{sec:Gillespie} a stochastic scheme based
on these transfer coefficients in terms of the Gillespie algorithm is
introduced. In Section \ref{sec:ME} a complementary master equation
scheme is described. In Section \ref{sec:Results} we apply our two
complementary formalisms to (i) the experimental constructs in
Ref. \cite{altan}; (ii) the T7 phage promoter sequence.  (iii) We show
a strong dependence on sequence, temperature and salt concentration
and demonstrate the good potential for nanosensing applications.
Technical details necessary for the introduction of our model appear in
a separate publication \cite{jcp}.

\section{General model and transfer rates}
\label{sec:General model}

\begin{figure}
\includegraphics[width=8.6cm]{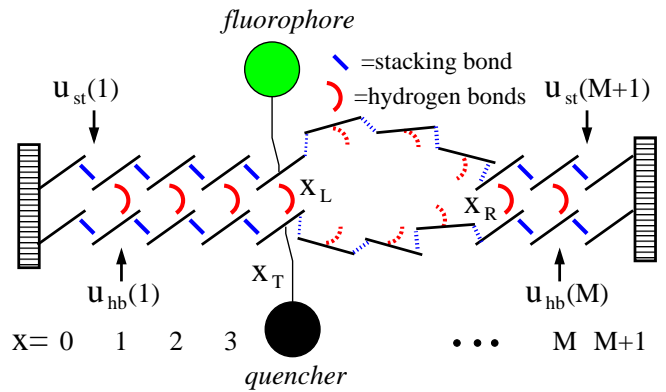}
\caption{Clamped DNA domain with internal bps $x=1$ to $M$, and tag position
$x_T$. The DNA sequence enters through the statistical weights $u_{\rm
st}(x)$ and $u_{\rm hb}(x)$ for disrupting stacking and hydrogen bonds
respectively. The bubble breathing process consists of the initiation
of a bubble and the subsequent motion of the forks at positions $x_L$
and $x_R$.
\label{fig:bubbles}}
\end{figure}

With typical experimental setups \cite{altan} in mind, we consider a
segment of double-stranded DNA with $M$ internal bps, that are clamped
at both ends, i.e., the bps $x=0$ and $x=M+1$ are always closed
(Fig.~\ref{fig:bubbles}). The heteropolymer character of
the problem enters via the position-dependence of the statistical
weights $u_{\rm hb}(x)=\exp\{\epsilon_{\rm hb}(x)/(k_BT)\}$ for
breaking the hydrogen-bonds of the bp at position $x$, and $u_{\rm
st}(x)=\exp\{\epsilon_{\rm st}(x)/ (k_BT)\}$ for disrupting the
stacking interactions between bps $x-1$ and $x$; $\epsilon_{\rm
st}(x)$ and $\epsilon_{\rm hb}(x)$ are the corresponding free
energies, which in general have energetic as well as entropic
contributions. Due to the high free energy barrier for bubble
initiation ($\xi\ll 1$, see below), opening and merging of multiple
bubbles are rare events,
such that a one-bubble description is appropriate \cite{tobiaslong}. The
positions $x_L$ and $x_R$ of the zipper forks respectively
correspond to the right- and leftmost closed bp of the bubble; these
are stochastic variables whose time evolution characterizes the bubble
dynamics. Note that writing the Boltzmann factors for the free energies
as $\exp\left\{\Delta G/(k_BT)\right\}$, a positive $\Delta G$
denotes an unstable bond.

In terms of $x_L$ and bubble size
$m=x_R-x_L-1$, the bubble partition factor is ($m\ge 1$)
\be
\mathscr{Z}(x_L,m)=\frac{\xi'}{(1+m)^c}
\prod_{x=x_L+1}^{x_L+m} \hspace*{-0.2cm}u_{\rm hb}(x)
\prod_{x=x_L+1}^{x_L+m+1}\hspace*{-0.2cm}u_{\rm st}(x),
\label{part}
\ee
completed by $\mathscr{Z}(m=0)=1$. Here, $\xi'=2^c\xi$, where
$\xi\approx 10^{-3}$ is the ring factor for bubble initiation from
Ref.~\cite{FK}. For a homopolymer $\xi$ is related to the
cooperativity parameter $\sigma_0\approx 10^{-5}$ \cite{poland,blake}
by $\sigma_0=\xi \exp\{\epsilon_{\mathrm{st}}/(k_BT)\}$ \cite{FK}. For
the entropy loss on forming a closed polymer loop we assign the factor
$(1+m)^{-c}$ \cite{blake,fixman} and take $c=1.76$ for the critical
exponent \cite{richard}. Note that a bubble with $m$ open bps needs
breaking of $m$ hydrogen bonds and $m+1$ stacking interactions, see
Eq.~(\ref{part}). The equilibrium probability for finding a bubble with
a given $x_L$ and $m$ is
  \be
P^{\rm eq}(x_L,m)=\frac{\mathscr{Z}(x_L,m)}{\mathscr{Z}(0)+\sum_{m=1}^M
\sum_{x_L=0}^{M-m}\mathscr{Z}(x_L,m)}.
\label{eq:P_eq}
  \ee
Below we impose the detailed balance condition when introducing the rates
to guarantee that $P_{\rm eq}(x_L,m)$ is indeed reached for long times.

Let us proceed by introducing the transfer (rate) coefficients for the
bubble dynamics. For the left zipper fork we define
$t^+_L(x_L,m)$ as the transfer coefficient for the process
$x_L\rightarrow x_L+1$, corresponding to bubble size decrease, and
$t^-_L(x_L,m)$ as the transfer coefficient for $x_L\rightarrow x_L-1$
(bubble size increase). For the right zipper fork we similarly
introduce $t^+_R(x_L,m)$ for $x_R\rightarrow x_R+1$ (bubble size
increase) and $t^-_R(x_L,m)$ for $x_R\rightarrow x_R-1$ (bubble size
decrease). In addition for the transition $m=0\rightarrow m=1$,
i.e., for the initial bubble opening process occurring at position $x_L$, we
introduce $t_G^+(x_L)$, and for the bubble closing process
$m=1\rightarrow m=0$ we employ $t_G^-(x_L)$. Note that $t_G^+(x_L)$
and $t_G^-(x_L)$ correspond to closing or opening of the bubble at
position $x=x_L+1$. Due to the clamping we require that $x_L\ge 0$ and
$x_R\le M+1$, and we therefore introduce reflecting conditions
  \be
t_L^-(x_L=0,m)=t_R^+(x_L,m=M-x_L)=0
\label{eq:reflecting1}
  \ee  
(also, $\mathsf{t}_L^+(x_L=-1,m)=0$ and $\mathsf{t}_R^-
(x_L,m=M-x_L+1)=0$ for $m=2,...,M+1$ for completeness).

Let us consider explicit forms for the transfer coefficients.
For bubble size decrease we take
  \be
t^+_L(x_L,m)|_{m\ge 2}=t^-_R(x_L,m)|_{m\ge 2}=
\mathcal{K}(m)/2\label{eq:t_L_plus}
  \ee
for the left fork and right forks, respectively.
We above defined the $m$-dependent rate coefficient
  \be
\mathcal{K}(m)=k m^{-\mu}.
\label{eq:rate_coeff}
  \ee
As in previous studies, this expression imposes the hook exponent
$\mu$, related to the fact that during the zipping process not only
the bp at the zipper fork is moved, but also part of the vicinal
single-strand is dragged or pushed
along \cite{Di_Marzio_Guttman_Hoffman,PRL,tobiaslong}. One would expect that
the hook exponent is only relevant for larger bubbles, and we put $\mu=0$
in the remainder of this work, mainly focusing on $T$ well below $T_m$, where
the bubbles sizes are small. The rate $k$
characterizes a single bp zipping. Its independence of $x$
corresponds to the view that bp closure requires the diffusional encounter
of the two bases and subsequent bond formation; as sterically AT and GC
bps are very similar, $k$ should not significantly vary with bp
stacking. $k$ is the only adjustable parameter of our model, and has
to be determined from experiment or future MD simulations. The factor
$1/2$ is introduced for consistency with previous approaches
\cite{tobias,tobiaslong}. We note that, in principle, an $x$-dependence of $k$
can easily be introduced in our approach by choosing different powers
of the statistical weights entering the rate coefficients such that
they still fulfill detailed balance.

 Bubble size increase is controlled by
\begin{eqnarray}
\nonumber
\mathsf{t}_{L}^{-}(x_L,m)&=& \mathcal{K}(m+1) u_{\rm st}(x_L) u_{\rm hb}(x_{L})
s(m)/2,\\
\mathsf{t}_{R}^{+}(x_L,m)&=&\mathcal{K}(m+1) u_{\rm st}(x_R+1) u_{\rm hb}(x_{R})
s(m)/2,
\label{eq:t_L_minus_etc}
\end{eqnarray}
for $m\ge 1$, where
\be
s(m)=\{(1+m)/(2+m)\}^c.
\ee
For $m\ge 1$ we thus
take the rate coefficients for bubble increase proportional to the
Arrhenius factor $u_{\rm st}u_{\rm hb}=\exp\{
(\epsilon_{\mathrm{hb}}+\epsilon_{\mathrm{st}})/[k_BT] \}$ multiplied
by the loop correction $s(m)$. Note that an unzipping event on
average involves the motion of one more open base-pair compared to a
zipping event, and the transfer coefficients above are therefore
proportional to $\mathcal{K}(m+1)$. Finally, bubble initiation and
annihilation from and to the zero-bubble ground state, $m=0
\leftrightarrow 1$ occur with rates
\bea
\nonumber
\mathsf{t}_G^+(x_L)&=& k \xi's(0)u_{\rm st}(x_L+1)u_{\rm hb}(x_L+1)u_{\rm
st}(x_L+2)\\
\mathsf{t}_G^-(x_L)&=&k.
\label{eq:t_G_plus}
\eea
with the bubble initiation factor $\xi'$ included in the expression
for $\mathsf{t}_G^+$. Note that $\mathsf{t}_G^+$, in contrast to the
opening rates for $m\ge1$, is proportional to an Arrhenius-factor
involving {\em two} units of stacking free energy. The annihilation rate
$\mathsf{t}_G^-(x_L)$ is twice the zipping rate of a single fork,
since the last open bp can close either from the left or right.  The
rates $\mathsf{t}$ together with the boundary conditions fully
determine the bubble dynamics.

We see that the rates $\mathsf{t}_L^{\pm}$, $\mathsf{t}_R^{\pm}$, and
$\mathsf{t}_G^{\pm}$ are chosen such that they fulfill the
detailed balance conditions:
\bea
&&\mathsf{t}_L^+(x_L-1,m+1) P_{\rm eq}(x_L-1,m+1)\nonumber\\
&&\hspace*{1.2cm}= \mathsf{t}_L^-(x_L,m)P_{\rm eq}(x_L,m),\nonumber\\
&&\mathsf{t}_R^-(x_L,m+1)P_{\rm eq}(x_L,m+1)\nonumber\\
&&\hspace*{1.2cm}=\mathsf{t}_R^+(x_L,m)P_{\rm eq}(x_L,m),\nonumber\\
&&\mathsf{t}_G^+(x_L)P_{\rm eq}(0)=\mathsf{t}_G^-(x_L)P_{\rm eq}
(x_L,1).\hspace*{0.2cm}
\label{eq:det_balance}
\eea
These conditions guarantee relaxation towards the equilibrium
distribution $P^{\rm eq}(x_L,m)$, see Eq.~(\ref{eq:P_eq}). In
the next two sections we use the above explicit expressions for the
transfer coefficients and describe the DNA breathing dynamics pursuing two
complementary approaches:
the stochastic Gillespie scheme (Section \ref{sec:Gillespie}) and 
the master equation (Section \ref{sec:ME}).

\section{Dynamic approaches to DNA-breathing}

\subsection{Gillespie approach}\label{sec:Gillespie}

In this section we use the Gillespie algorithm together with the
explicit expressions for the transfer coefficients introduced in the
previous section in order to generate sequence specific stochastic time
series of breathing bubbles. In particular we show how the motion of a
tagged bp is obtained.

To denote a bubble state of $m$ broken bps at position $x_L$ we
define the occupation number $b(x_L,m)$
with the properties $b(x_L,m)=1$ if the particular
state $\{x_L,m\}$ is occupied and $b(x_L,m)=0$ for unoccupied states. For
the completely zipped state $m=0$ there is no dependence on $x_L$, and we
introduce the occupation number $b(0)$. The stochastic DNA
breathing then corresponds to the nearest neighbor jump
processes in the lattice of permitted states \cite{jcp}. In the Gillespie
scheme, each jump away from the state $\{x_L,m\}$ (i.e., from the state with
$b(x_L,m)=1$) occurs at a random time $\tau$, and in a random
direction to one of the nearest neighbor states. This stochastic process 
is governed by the reaction probability density function
\cite{suman,Gillespie,Gillespie1}
\begin{equation}
\label{gill}
P(\tau,\mu,\nu)=\mathsf{t}^{\mu}_{\nu}(x_L,m)\exp\left(-\tau\sum_{\mu,\nu}
\mathsf{t}^{\mu}_{\nu}(x_L,m)\right).
\end{equation}
More explicitly, for a given state $(x_L,m)$ the joint probability density
(\ref{gill}) defines after what waiting time $\tau$ after the previous
random step the next step occurs, and in which reaction pathway, $\nu\in\{
G,L,R\}$, $\mu\in \{+/-\}$. In the present case, $\nu$ and $\mu$ denote 
$x$-dependent  zipping or unzipping of a bp at the left of right zipper fork.
A simulation run produces a time series of occupied
states $\{x_L,m\}$ and how long time $\tau=\tau_j$ ($j=1,...,N$, where
$N$ is the number of steps in the simulation) this particular state is
occupied. This waiting time $\tau$, in particular, according to
Eq.~(\ref{gill}) follows a Poisson distribution \cite{cox}.
Note that the waiting times governed by (\ref{gill}) vary widely, as the
reaction rates occur in the exponential (in particular, the bubble initiation
with the $\xi$-factor has a long characteristic time scale). The
fact that the Gillespie scheme uses the weighted reaction time scale instead
of fixed simulation time steps makes this algorithm very efficient.

\subsubsection{Tagged bp survival and waiting time densities}

Motivated by the experimental setup in \cite{altan} we study the
motion of a tagged bp at $x=x_T$, see Fig.~\ref{fig:bubbles}. In the
fluorescence correlation experiment fluorescence occurs if the bps in a
$\Delta$-neighborhood of the fluorophore position $x_T$ are open
\cite{altan}. Measured fluorescence time series thus correspond to the
stochastic variable $I(t)$, with the properties
$I(t)=1$ if at least all bps in $[x_T-\Delta,x_T+\Delta]$ are open,
and $I(t)=0$ otherwise. Thus, if $I=1$ we are in the phase space region
defined by 
  \be
\mathbb{R}1:\{ 0 \leqslant x_L \leqslant x_T-\Delta-1, \ x_T-x_L+\Delta
\leqslant m \leqslant M-x_L \}.
  \ee
Conversely, $I=0$ corresponds to the complement $\mathbb{R}0$ of
$\mathbb{R}1$. The stochastic
variable $I(t)$ is then obtained by summing the Gillespie occupation
number $b(x_L,m)$ ($b(x_L,m)$ takes only values $0$ or $1$) over
region $\mathbb{R}1$, i.e.,
  \be
I(t) =  \sum_{x_L,m\in \mathbb{R}1} b(x_L,m).
  \ee
From the time series for $I(t)$ one can, for instance, calculate the
waiting time distribution $\psi(\tau)$ of times spent in the $I=0$
state, as well as the survival time distribution $\phi(\tau)$ of times
in the $I=1$ state. Explicit examples for $\psi(\tau)$ and
$\phi(\tau)$ are shown in Section \ref{sec:Results}.

The probability that the tagged bp is
open becomes 
  \be
P_G(t_j)=\frac{1}{t_N}\sum_{j=1}^{N} \tau_j I(t_j),
  \ee
where $t_j=\sum_{j'=1}^{j} \tau_{j'} $. For long times the explicit
construction of the Gillespie scheme together with the detailed
balance conditions guarantee that $P_G(t_j)$ tends to the equilibrium
probability, i.e., that $P_G(t_j\rightarrow\infty)=\sum_{x_L,m\in \mathbb{R}1}
P^{\rm eq}(x_L,m)$, where $P^{\rm eq}(x_L,m)$ is given in
Eq.~(\ref{eq:P_eq}).

\subsubsection{Tagged base-pair autocorrelation function}

The autocorrelation function for a tagged bp is obtained through 
\begin{eqnarray}
\nonumber
A_t(x_T,t)&=&\overline{I(t)I(0)}-(\overline{I(t)})^2\\
&&\hspace*{-1.8cm}
=\frac{1}{T}\int_0^TI(t+t')I(t')dt'-\left(\frac{1}{T}\int_0^TI(t')dt'\right)^2
\label{eq:A_t_xT}
\end{eqnarray}
which for long sampling times $T$ converges to the ensemble average,
Eq.~(\ref{eq:A_t}), from the master equation (introduced in the next
section). The function $A_t(x_T,t)$ corresponds to the quantity
obtained in the fluorescence correlation experiment of Ref.~\cite{altan}.

\subsection{Master equation formulation}
\label{sec:ME}

Complementary to the stochastic simulations of DNA-breathing detailed
in the preceding
section we here introduce a master equation for the joint probability density
$P(t)=P(x_L,m,t;x_L',m',0)$ that at time $t$ the system is in state $\{x_L,m\}$
and that it was in state $\{x_L',m'\}$ at $t=0$. The master equation, which
is equivalent (in the sense
that it produces the same averaged quantities) to the Gillespie scheme, can
be formally written as
  \be
\frac{\partial}{\partial t}P(t)= \mathscr{W} P(t),
\label{eq:master_eq_main}
  \ee
where the explicit form of the matrix $\mathscr{W}$ is given in terms
of the rate coefficients from the previous section in Ref.~\cite{jcp}.
A standard approach to the master equation is
the spectral decomposition \cite{van_Kampen,Risken}
  \be
P(t)=\sum_p
c_p Q_p \exp(-\eta_p t).
\label{eq:spectral_decomp}
  \ee 
The
coefficients $c_p$ are obtained from the initial condition. Inserting
Eq.~(\ref{eq:spectral_decomp}) into Eq.~(\ref{eq:master_eq_main}) produces
the eigenvalue equation
  \be
 \mathscr{W} Q_p=-\eta_p Q_p.
 \label{eq:eigenvalue_eq}
  \ee
From the eigenvalues $\eta_p$ and eigenvectors $Q_p$ of
Eqs.~(\ref{eq:eigenvalue_eq}), any quantity of interest can be
constructed.

\subsection{Dynamic quantities for a tagged bp}
\label{sec:ME_surv_wait}

The waiting time density $\psi(t)$ and the survival time density
$\phi(t)$, as obtained in a Gillespie scheme, correspond to the first passage
problem to start from an initial state with $I=1$ ($I=0$) and passing to
$I=0$ ($I=1$). It is discussed in detail in Ref.~\cite{jcp}
how these quantities can be obtained from the master
Eq.~(\ref{eq:master_eq_main}).

The equilibrium autocorrelation function
\be
A(x_T,t)=\langle I(t)I(0)\rangle -(\langle I \rangle) ^2
\label{eq:A_t}
\ee
is a measure for the relaxation dynamics of the tagged bp. This can be
seen from the identity
\be
\langle I(t)I(0)\rangle=\sum_{I=0}^1 \sum_{I'=0}^1 I \rho(I,t;I',0)I'=
\rho(1,t;1,0),
\label{eq:I_t}
\ee
where $\rho(1,t;1,0)$ is the survival probability density
that $I(t)=1$ and that $I(0)=1$ 
initially. Using the fact that $\rho(1,t;1,0)$ is obtained by summing
$ P(x_L,m,t;x_L',m',0)$ exclusively over region $\mathbb{R}1$ we obtain
\be
\rho(1,t;1,0)=\sum_{x_L,m,x_L',m'\in \mathbb{R}1} P(x_L,m,t;x_L',m',0).
\ee
Combining this result with Eq.~(\ref{eq:I_t}), the spectral decomposition
(\ref{eq:spectral_decomp}), and assuming that we initially are at
equilibrium:
$P(x_L,m,0;x_L',m',0)=\delta_{mm'}\delta_{x_Lx_L'}P_{\rm eq}(x_L,m)$, the
autocorrelation function (\ref{eq:A_t}) can be rewritten as
  \be
A(x_T,t)=\sum_{p\neq 0} \left[T_p(x_T)\right]^2
\exp(-t/\tau_p),\label{eq:A_xT_main}
  \ee
with relaxation times $\tau_p=1/\eta_p$, and where 
\be
T_p(x_T)=\sum_{x_L=0}^{x_T-\Delta-1} \sum_{m=x_T-x_L+\Delta}^{M-x_L} Q_p(x_L,m).
\label{eq:T_p}
\ee
For long times, i.e., when the time average is long enough, $A(x_T,t)$
agrees with $A_t(x_T,t)$ given in Eq.~(\ref{eq:A_t_xT}) as will be
illustrated in the next section.  We can rewrite the correlation
function according to the spectral decomposition
\be
A(x_T,t)=\int d\tau \exp(-t/\tau) f(x_T,\tau),
\ee
where we introduced the weighted spectral density
\be
f(x_T,\tau)=\sum_{p\neq 0} [T_p(x_T)]^2 \delta(\tau-\tau_p). \label{eq:f_tau}
\ee
This relaxation time spectrum directly provides the spectral content
of the relaxation behavior of the DNA-bubble, and sometimes
a better (but equivalent) visualization of the system than the
autocorrelation function.

\section{Results}
\label{sec:Results}

In this section we apply our two complementary formalisms to study the
behavior of (i) the designed DNA constructs used in the experiments
of Ref.~\cite{altan}, and (ii) the T7 phage promoter sequence.

\subsection{Comparison to experimental results}\label{sec:experiments}

\begin{figure}
\includegraphics[width=8.6cm]{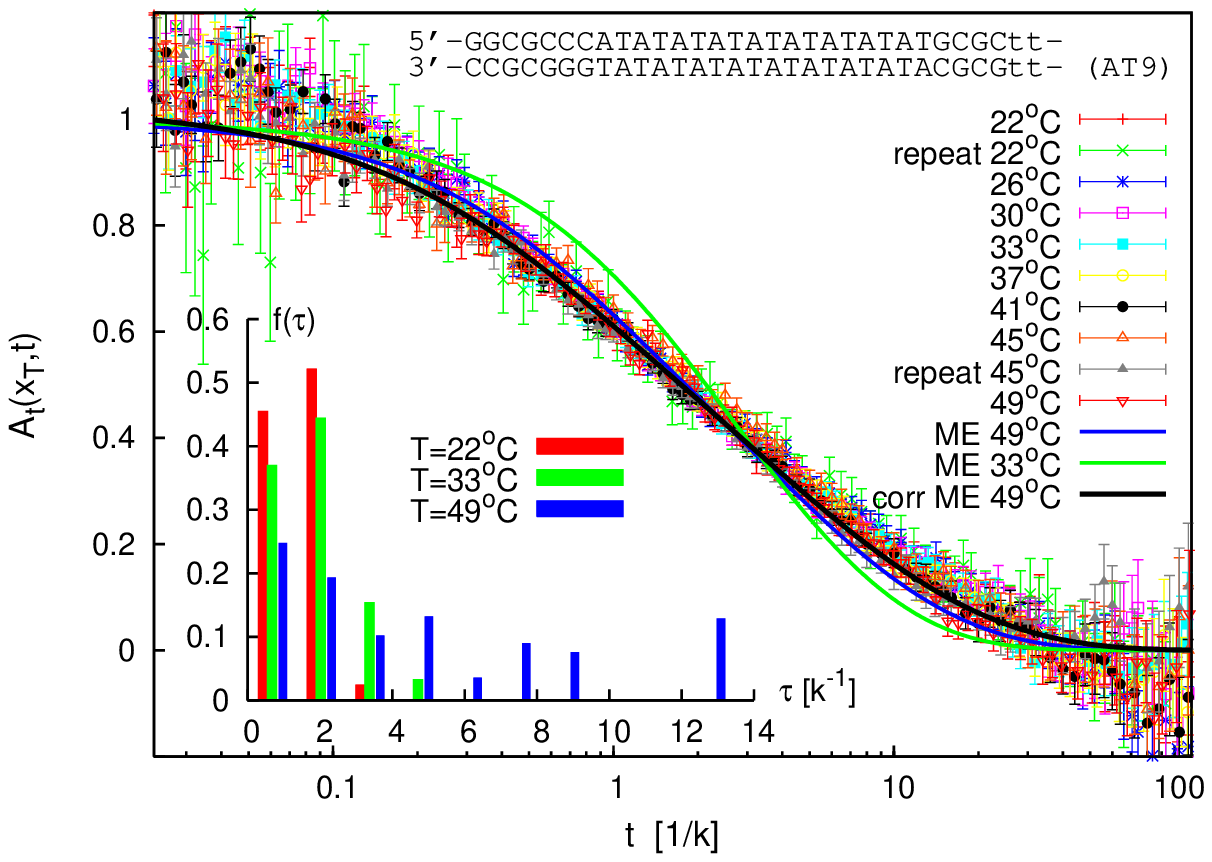}
\includegraphics[width=8.6cm]{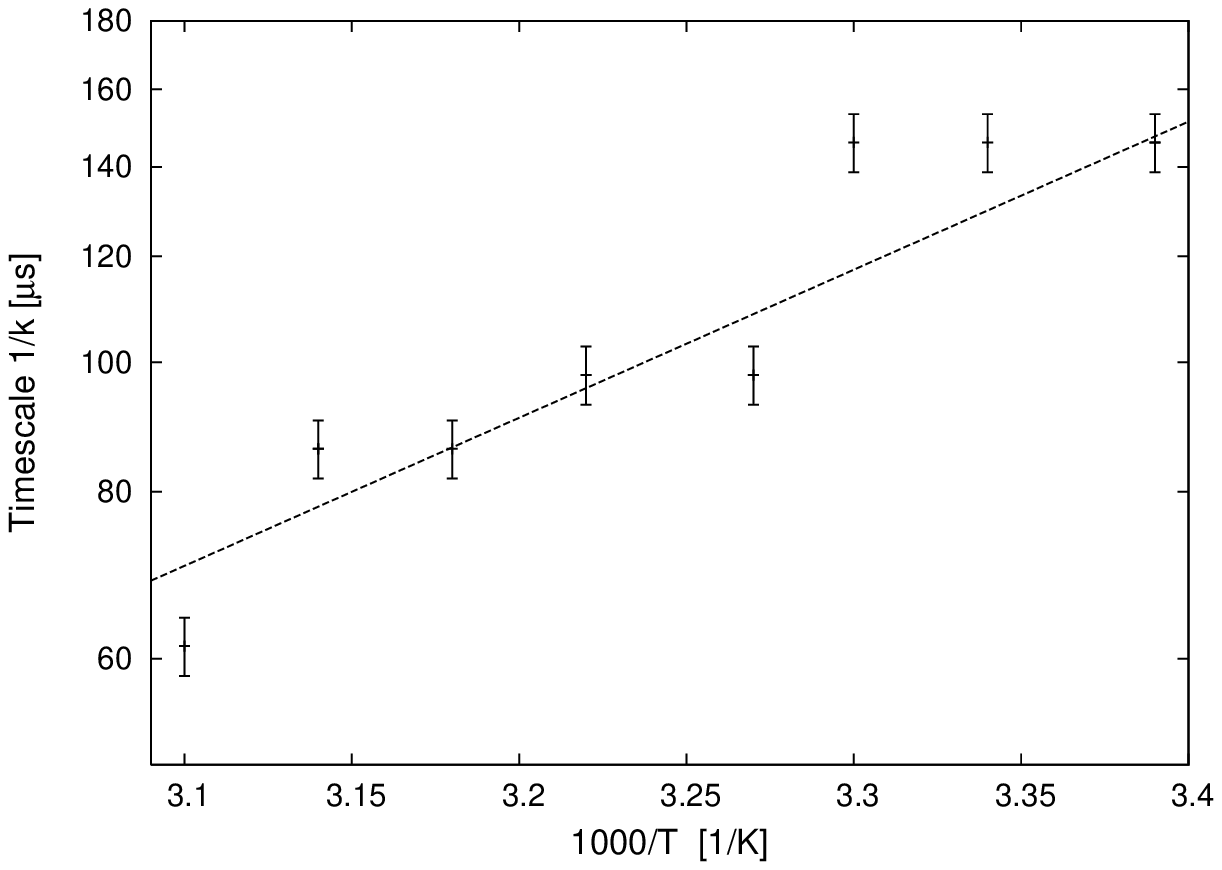}
\caption{Top: Autocorrelation function $A_t(x_T,t)$ at various
temperatures $T$ measured for the sequence AT9 from \cite{altan} at
100 mM NaCl. The sequence is indicated in the figure, where the four
lower case t's symbolize a small bulge loop. The full lines show the
results from the master equation based on the DNA parameters from \cite{FK}.
Inset: Relaxation time spectrum $f(\tau)=f(x_T,\tau)$, showing broadening
with increasing temperature. Bottom: Characteristic zipping time $1/k$
as a function of temperature in an Arrhenius plot. The line shows a least
squares fit to an Arrhenius law $\tau\propto\exp(A/T)$ with $A=2.6\times
10^{3}$K.}
\label{autocorr}
\end{figure}

\begin{figure}
\includegraphics[width=8.6cm]{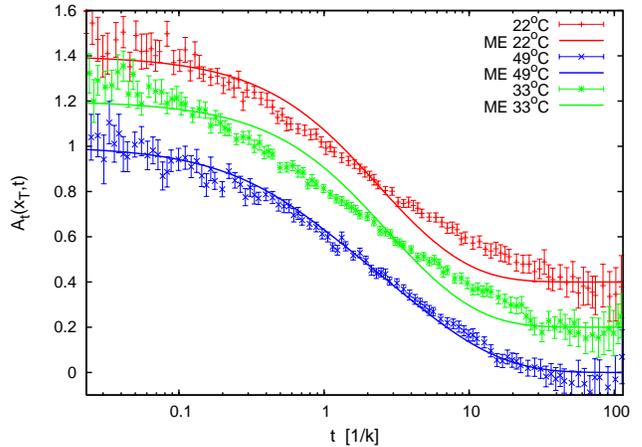}
\caption{Individual autocorrelation functions $A_t(x_T,t)$ for
three different temperatures (22$^\circ$C, 33$^\circ$C, 49$^\circ$C)
spanning the $T$-range probed in the fluorescence experiment. While for
the highest temperature, the match between data and theory is very good,
deviations occur at lower temperatures. Reasons for these deviations are
discussed in the text. Note that the data for 33$^\circ$C are shifted
vertically by 0.2, the ones for 22$^\circ$C by 0.4.}
\label{autocorr_45}
\end{figure}

In Fig.~\ref{autocorr} the autocorrelation functions $A_t(x_T,t)$ for
the sequence AT9 from \cite{altan} are shown for various temperatures
$T$. The data were scaled by $k$ such that the curves coincide where
$A(t)=1/2$.
The strong scatter at short times is mainly ascribed to quantum transitions
in the fluorophore \cite{altan,oleg1}. The
lower graph shows the temperature dependence of the characteristic
zipping time, $1/k$. Individual autocorrelations for three temperatures
are compared in Fig.~\ref{autocorr_45}.

In the combined autocorrelation plot, Fig.~\ref{autocorr}, the
black line shows the predicted behavior of $A(x_T,t)$, calculated by
numerical solution of the eigenvalue Eq.~(\ref{eq:eigenvalue_eq})
by help of Eq.~(\ref{eq:A_xT_main}). Stability parameters from \cite{FK}
for $T=49^\circ$C and 100 mM NaCl concentration were used. As in the
experiment we assumed that fluorophore and quencher attach to bps
$x_T=17$ and $x_T+1$, and that both are required open to produce a
fluorescence signal (the outermost GC-pairs in the sequence given in figure
\ref{autocorr} were taken as clamped, i.e., labeled as $x=0$ and $x=M+1$).
From the scaling plot, we calibrate the zipping
rate as $k=7.1\times 10^4/$s for $T=49^\circ$ in good agreement with
the findings from Ref.~\cite{altan}. The calculated behavior
reproduces the data within the error bars. The green curve corresponds to the
ME result for $T=33^\circ$C, showing more pronounced deviations from
experimental data. Notice that for lower temperatures the
relaxation time distribution $f(x_T,\tau)$ becomes narrower (Fig.~\ref{autocorr}
inset). Thus, our model predicts that the dynamics for smaller
temperatures involve fewer modes, which is in contrast to the experimental data
that have a broad, multimodal behavior also for low temperatures.

The individual behavior of the autocorrelation is dissected in
Fig.~\ref{autocorr_45} for three temperatures spanning the full $T$-range
probed in the fluorescence experiments. Note the good quality of the match
between experimental data and model prediction for the highest temperature
(49$^\circ$C). This temperature is already close
to the denaturation temperature of the bubble domain of the AT9 construct
(the contribution of the longest relaxation time in the rather broad
spectrum of relaxation times is considerably larger than the three previous
ones). The tendency of overestimation of the slope in the autocorrelation
function by our model at lower temperatures is obvious for curves at
22$^\circ$C and 33 $^\circ$C, while the experimental slope remains almost
constant over this $T$-range.

We expect three effects to contribute to the deviations by broadening the
relaxation time spectrum, i.e., lowering the free energy of the system:

(i) In the present fluorescence correlation spectroscopy experiments, two
contributions superimpose to produce the fluorescence signal \cite{oleg1}:
The diffusional motion of the molecule carrying the fluorophore in and out
of the confocal volume, and the actual breathing dynamics. Without the
breathing the autocorrelation function takes the
form $A(t)\sim 1/(1+t/\tau_D)$ (for a narrow beam waist), where
$\tau_D=w^2/(4D)\approx 150$ms, with $w$ being the linear size of the beam
waist and $D$ is the diffusion constant of the construct. In \cite{altan}
the pure diffusive contribution was eliminated by performing a separate
experiments with the quencher being removed (measuring the solely diffusive
contribution), and dividing out this result from the signal. However,
as the quencher is removed the diffusion constant of the construct is
slightly changed. In order to roughly account for this fact, the blue
curve shown in Fig.~\ref{autocorr} was obtained by a 3\% reduction of
the diffusion time $\tau_D$. Note that the agreement of the blue line
with the data is excellent. This underlines the sensitivity of the
DNA-breathing single molecule data, pointing toward potential dynamic
methods to calibrate both $k$ and $\Delta G$.

(ii) It is very likely that the presence of the fluorophore and quencher
molecules destabilizes the DNA---despite the short stalk through which
fluorophore and quencher are attached---by altering the Watson-Crick
and stacking interactions. The resulting decrease of the stacking free
energy therefore is expected to effect a lower free stacking energy
in comparison to the undressed DNA, for
which the stability data are measured and which are used in our model.

\begin{figure}
\includegraphics[width=8.6cm]{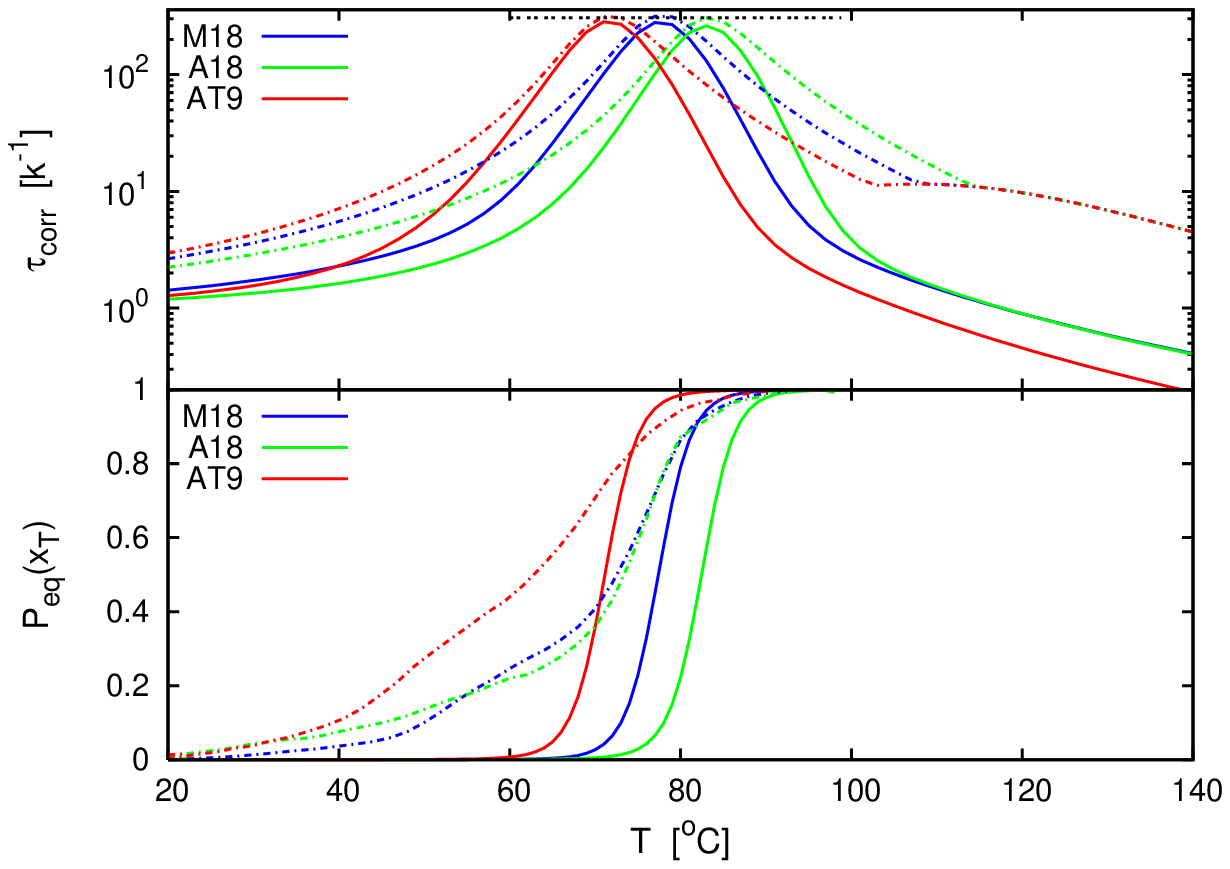}
\caption{Top: Mean correlation time versus versus temperature for the
three constructs in reference \cite{altan}. The solid curves are for
the mean correlation time $\tau_{\rm corr}$, whereas the dashed curves
are the longest relaxation time $\tau_1$. Bottom: Melting curves. The
dashed curves are experimental results. For the theoretical curves we
used parameters from \cite{FK} for 100 mM NaCl concentration.}
\label{tau_temp}
\end{figure}

(iii) Finally, our present model does not take into account the entropic
contributions due to the degrees of freedom of the fluorophore/quencher pair,
i.e., the fact that for bigger bubbles the fluorophore/quencher pair has more
freedom to diffuse around and rotate. To approximately account for this we
would change the partition from $\mathscr{Z}(x_L,m)$ to $\Omega_{\mathrm{FQ}
}(x_L,m)\mathscr{Z}(x_L,m)$, where $\Omega_{\mathrm{FQ}}(x_L,m)$ is the number
of configurations for the fluorophore/quencher pair for a given bubble size
and position. To demonstrate this effect assume for simplicity that each bps
that opens up provides one unit of entropy, $\Delta S_{\mathrm{FQ}}$, so that
$\Omega_{\mathrm{FQ}}=e^{m \Delta S_{\mathrm{FQ}}/k_B}$, i.e., effectively we
increase the statistical weight $u$ according to $u\to ue^{\Delta S_{\mathrm{
FQ}}/k_B}$, leading to a shift in the melting curve towards lower temperatures
(as seen in experiments, compare also Fig.~\ref{tau_temp}). In reality,
however, we would expect a more intricate $m$-dependence of the complexions
$\Omega_{\mathrm{FQ}}$, for instance, it may be so that as the first bps
close the the tag-position open up a relatively large amount of
fluorophore/quencher entropy is released, while further opening of bps 
contributes less.

We stress once more that all three effects will broaden the relaxation
time spectrum.
Further control experiments will be needed to obtain more precise
information on the effects caused by the presence of the fluorophore
and quencher molecules.

The activation plot in Fig.~\ref{autocorr} was obtained from the construction
of the scaling plot for the blinking autocorrelation function by the relative
shift of the individual curves along the logarithmic time axis. The
corresponding error bars were estimated from the width of the collapsed data
at the midpoint ($A_t(x_T,t)=1/2$) as 20 per cent of the absolute value. The
real experimental error is likely to be higher.
However, it is difficult to estimate. The activation plot indicates an
Arrhenius-type behaviour, that is probably due to an energetic barrier
crossing when the bp-bond establishes.

We point out that we here only considered the AT9 sequence from
\cite{altan}, and not the other to constructs A18 and M18. The latter
two constructs have 4 or more consecutive AT-bps, and it is known that
such sequences assume the B'-conformation rather than the usual
$B$-structure \cite{russu} for which the parameters of \cite{FK}
apply. In B' DNA, the breathing dynamics is significantly altered
\cite{russu}. Fitting our model to the A18 and M18 constructs, we found indeed
that these sequences showed more pronounced deviations from our model.

In Fig. \ref{tau_temp} the top panel show the mean correlation time
$\tau_{\rm corr}\equiv \int_0^\infty \tau f(x_T,\tau) d\tau
=\int_0^\infty A(x_T,t)dt$, see Eqs.~(\ref{eq:A_xT_main}) and
(\ref{eq:f_tau}), for the three constructs of Ref. \cite{altan}; these
constructs all consist of 18 consecutive AT-bps with end-clamps
consisting of GC-pairs. The bottom
panel depicts the probability $P_{\rm eq}(x_T)$ that the bps at $x_T$
and $x_T+1$ are open, i.e., the probability to get a fluorescence
signal. We notice that $\tau_{\mathrm{corr}}$ has pronounced maxima
at the melting transition (the point where $P_{\rm eq}=1/2$ in the
bottom panel). This critical slowing down at the melting is indeed a
characteristic signature of a phase transition, compare Ref.~\cite{bicout}.
Notice that the experimental results (dashed lines) for
$P_{\rm eq}(x_T)$ deviate from the one predicted here, indicating
that the fluorophore-quencher pair indeed has a destabilizing effect
on the DNA helix. Also note the different melting behaviors of the
three construct despite identical AT and GC contents predicted here as
well as by experiments; this illustrates the importance of stacking
interactions. Also notice that
there is nice agreement between our theoretical results and
experiments concerning the relative ordering of the melting
temperatures: AT9 melts first, and A18 last. The horizontal line
($\tau_{\mathrm{max}}$ 1D) in the top panel represents the longest
relaxation time $(2M+1)^2/ \pi^2 k^{-1}$ obtained from the homopolymer
model of Ref.~\cite{tobias,tobiaslong} in the limit $u\to 1$, $\sigma_0\to0$
and $c=0$ (for $M=27$, length of the three constructs), thus giving a
scaling consistent with the first exit of unbiased diffusion, see
Ref.~\cite{jcp}.

\subsection{Bacteriophage T7}\label{sec:T7}

\begin{figure}
\includegraphics[width=8.6cm]{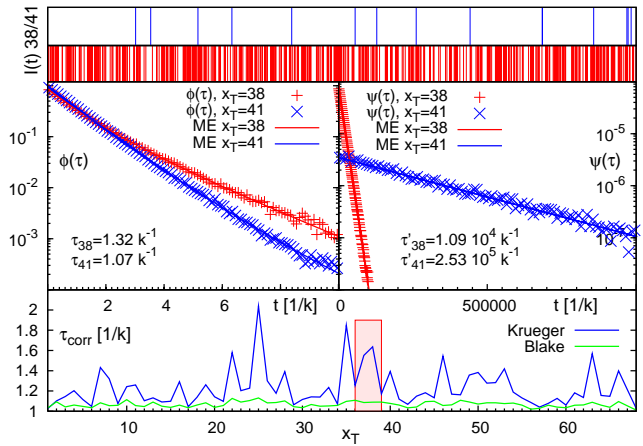}
\caption{Top: Fluorescence time series $I(t)$ for the T7 promoter
sequence, with tag position $x_T=38$ (red) and $x_T= 41$
(blue). Middle: Waiting time ($\psi(\tau)$) and fluorescence survival time
($\phi(\tau)$) densities, in units of $k$. The data points (solid lines) are
results from the Gillespie algorithm (master equation). Bottom: Mean
correlation time for $\Delta=0$. All results are for $T=37
^\circ C$ and 100 mM NaCl with DNA parameters from \cite{FK}. }
\label{signal}
\end{figure}

By master equation and stochastic simulation we investigate the promoter
sequence of the T7 phage (a bacteriovirus). A promoter is a sequence
(often containing the 4 bp long TATA motif) marking the start of a gene,
to which RNA polymerase is recruited and where transcription then
initiates. Previous studies \cite{Kalosakas_Rasmussen,Kalosakas_Rasmussen1} based on the
Dauxois-Peyrard-Bishop model found that the the TATA motif is
characterized by a particularly low stability and therefore proneness to
bubble formation, although the statistical relevance of those data were
under discussion \cite{vanErp}.
We here revisit the problem of the stability and dynamics of the TATA motif
using the necessary full set of stacking interactions.
The T7 promoter sequence we investigate is
\begin{equation}
\begin{array}{l}
\mbox{\small\texttt{\textcolor{white}{AAAA}1\textcolor{white}{%
AAAAAAAAAAAAAAAAAA}20}}\\[-0.15cm]
\mbox{\small\texttt{\textcolor{white}{AAAA}|\textcolor{white}{%
AAAAAAAAAAAAAAAAAA}|\textcolor{white}{AAAAAAAAA}\textcolor{white}{AAA}}}\\[-0.15cm]
\mbox{\small\texttt{5'-aTGACCAGTTGAAGGACTGGAAGTAATACGACTC}}\\
\mbox{\small\texttt{\textcolor{white}{AAA}AG}\textcolor{red}{
\texttt{TATA}}\texttt{GGGACAATGCTTAAGGTCGCTCTCTAGGAg-3'}}\\[-0.15cm]
\mbox{\small\texttt{\textcolor{white}{AAAAAAA}|\textcolor{white}{AA}|
\textcolor{white}{AAAAAAAAAAAAAAAAAAAAAAAAA}|\textcolor{white}{AAA}}}\\[-0.15cm]
\mbox{\small\texttt{\textcolor{white}{AAAAAAA}\textcolor{red}{38}%
\textcolor{white}{A}\textcolor{blue}{41}\textcolor{white}{%
AAAAAAAAAAAAAAAAAAAAAAAAA}68\textcolor{white}{AAA}}}
\end{array}
\end{equation}
whose TATA motif is marked red
\cite{Kalosakas_Rasmussen,Kalosakas_Rasmussen1}. Fig.~\ref{signal} shows the time series of
$I(t)$ at $37^{\circ}$C for the tag positions $x_T=38$ in the core of
TATA, and $x_T=41$ at the second GC bp after TATA. Bubble events occur
much more frequently in TATA (the TA/AT stacking interaction is
particularly weak \cite{FK}). This is quantified by the density of
waiting times $\psi(\tau)$ spent in the $I(t)=0$ state, whose
characteristic time scale $\tau'=\int_0^\infty d\tilde{\tau}\tilde{\tau}
\psi\left(\tilde{\tau}\right)$
is more than an order of magnitude longer than at $x_T=41$. In contrast, we
observe similar behavior for the density of opening times $\phi(\tau)$ for
$x_T=38$ and $41$, where the characteristic time is
$\tau=\int_0^\infty d\tilde{\tau} \tilde{\tau} \phi(\tilde{\tau}) $.
The solid lines are the results from the master equation,
see subsection \ref{sec:ME_surv_wait}, showing excellent agreement with
the Gillespie results. Notice that whereas $\psi(t)$ is characterized
by a single exponential, $\phi(t)$ show a crossover between different
regimes. For long times both $\psi(\tau)$ and $\phi(\tau)$ decay
exponentially as they should for a finite DNA stretch. As shown in the
bottom for the parameters from \cite{FK}, the variation of the mean
correlation time $\tau_{\mathrm{corr}}=\int A(x_T,t)dt$ obtained from
the ME is small for the entire sequence, consistent with the low
sensitivity to the sequence of $\phi(\tau)$. However, note the even
smaller variation predicted for the parameters of \cite{blake},
indicating that the stability parameters of \cite{FK} are more
sequence sensitive compared to previously used values \cite{blake}.
We speculate that the recurrence frequency of bubble events may be 
a clue in the understanding of transcription initiation: If the protein,
that is supposed to bind to the specific site, senses a time-averaged
energy landscape, the significantly more frequent bubble events at TATA
may trigger its binding and thus trigger transcription initiation.

\begin{figure}
\includegraphics[width=8.6cm]{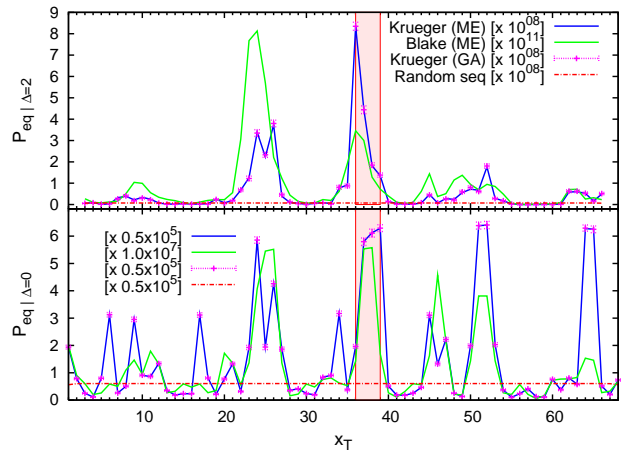}
\caption{Probability to have at least the bps $[x_T-\Delta,x_T+\Delta]$ open
(as assumed to be necessary for fluorescence to
occur) for different tag positions $x_T$, and for $\Delta=0$ and
$2$. The results from master equation results and Gillespie algorithm
show excellent agreement. Same parameters as in Fig. \ref{signal}.}
\label{open_symm}
\end{figure}

Fig.~\ref{open_symm} shows the equilibrium probability that the bps
$[x_T -\Delta,x_T+\Delta]$ are open, as necessary for fluorescence to
occur. We plot data obtained from the zeroth mode (an ME eigenvalue
problem always has one zero eigenvalue, the corresponding eigenvalue
is the equilibrium probability \cite{van_Kampen}) of the ME together
with the time average from the stochastic simulation (G), finding
excellent agreement.  Whereas for $\Delta=0$ several segments show
increased tendency to denaturation, for the case $\Delta=2$, one major
peak is observed; the data from \cite{FK} coincide precisely with
TATA. For comparison the equilibrium probability obtained using DNA stability
data from \cite{blake} have their maximum peak upstream. Analysis for various
$\Delta$ indicate best discrimination of the TATA sequence being open
for $\Delta=2$. Biologically, this finding is significant, as it
corresponds to the probability for simultaneous opening of the
whole TATA motif. For future FCS or energy transfer experiments investigating
the relevance of denaturation-induced facilitation of transcription
initiation, it
therefore appears important to optimize the $\Delta$-dependence for
best resolution, e.g., by adjusting the linker lengths of fluorophore
and quencher. This could, in principal, be experimentally achieved as,
assuming a circular bubble of 5 open bps with bp-bp distance 3.4{\AA},
the distance between fluorophore and quencher on bubble opening
increases by 6-7 {\AA}, the same magnitude as the F{\"o}rster transfer
radius.

In Fig.~\ref{open_symm} we compare the opening probabilities to the values
for a random sequence, for which we chose the free energies such that the
content of AT and GC bps is 50:50. Then, we define
\be
\epsilon_{\mathrm{hb,random}}=\epsilon_{\mathrm{hb,AT}}/2+\epsilon_{
\mathrm{hb,GC}}/2,
\ee
for the hydrogen bonding, and
\begin{eqnarray}
\nonumber
\epsilon_{\mathrm{st,random}}&=&\frac{1}{16}\left(\epsilon_{\mathrm{st,AT/TA}}+
\epsilon_{\mathrm{st,TA/AT}}+2\epsilon_{\mathrm{st,AT/AT}}\right.\\
\nonumber
&&\hspace*{-1.6cm}+\epsilon_{\mathrm{st,GC/CG}}+\epsilon_{\mathrm{st,CG/GC}}
+2\epsilon_{\mathrm{st,GC/GC}}+2\epsilon_{\mathrm{st,GA/CT}}\\
&&\hspace*{-1.6cm}\left.+2\epsilon_{\mathrm{st,CA/GT}}
+2\epsilon_{\mathrm{st,AG/TC}}+2\epsilon_{\mathrm{st,AC/TG}}\right)
\end{eqnarray}
for the stacking free energies. The numerical values are $\epsilon_{\mathrm{
hb,random}}=0.4$ kcal/mol and $\epsilon_{\mathrm{st,random}}=-1.6$ kcal/mol
at $T=37 ^\circ C$ and 100 mM NaCl. Inserting both into the expressions for
the partition factor (\ref{part}) and the rates $\mathsf{t}$, the dashed
lines in Fig.~\ref{open_symm} are obtained. Thus, the peaks in the opening
probabilities are distinctly significant.

\begin{figure}
\includegraphics[width=8.6cm]{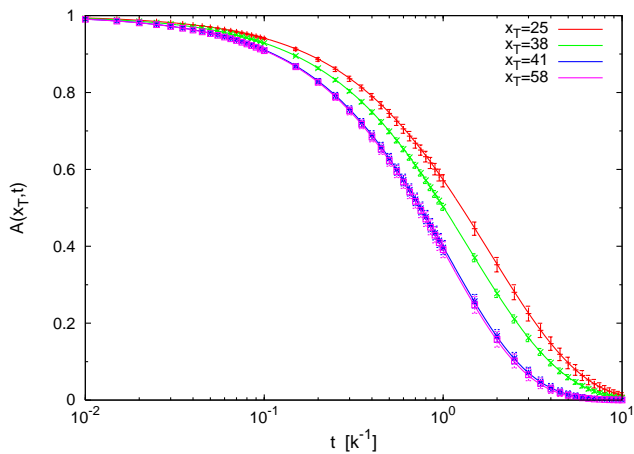}
\caption{Autocorrelation function $A(x_T,t)$ for a tagged bp for the
  T7 sequence. The solid lines are master equation results whereas the
  data points represent results from the Gillespie scheme. Note that
  the autocorrelation function is sensitive to the DNA sequence as
  well as the tagging position.  Same parameters as in
  Fig. \ref{signal}}
\label{fig:A_t_general}
\end{figure}

In order to illustrate the breathing dynamics of the T7 sequence using
experimentally measurable quantities, Fig.~\ref{fig:A_t_general} shows
the autocorrelation functions (see subsection \ref{sec:T7}), for four
different tag positions $x_T$ (same parameters as above) within the promoter
region. Both the Gillespie approach as well as the master equation were used
and compared; excellent agreement between them are found. The autocorrelation
function for the tagged bp decays faster if positioned in a GC-rich region
than in an AT-rich region. Comparing with Fig.~\ref{autocorr}
it should be possible to resolve the different decay times of the
autocorrelation function experimentally.

\subsection{Nanosensing applications}

\begin{figure}
\includegraphics[width=8.6cm]{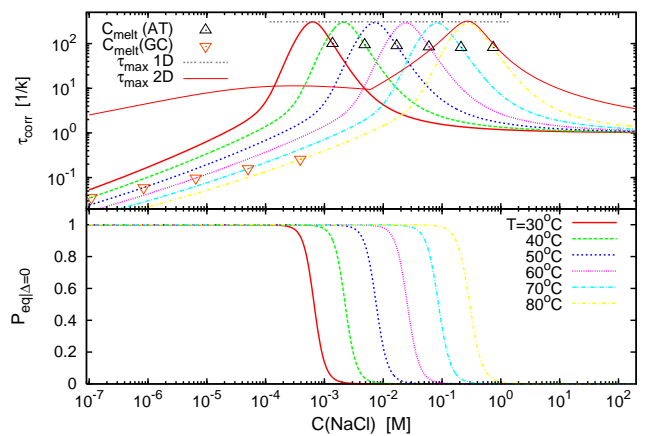}
\caption{Top: Mean correlation time versus NaCl concentration for
various temperatures $T$ for the AT9 construct, showing a critical
slowing down at the melting concentration (compare lower panel). The
triangles denotes the melting concentration for infinitely long random
AT and GC stretches. Bottom: Opening probability for $\Delta=0$.}
\label{salttemp}
\end{figure}

In Ref.~\cite{FK} the DNA Watson-Crick and stacking parameters were
obtained for different NaCl concentrations, allowing us to study
the effect of salt concentration on the breathing dynamics and equilibrium
properties of DNA. Fig.~\ref{salttemp} shows the dependence of the
mean correlation time $\tau_{\mathrm{corr}}$ and the equilibrium opening
probability $P_{\rm eq}(x_T)$ for the AT9 sequence on salt
concentration $C$ and temperature $T$, using the same tagging position as in
subsection \ref{sec:experiments}. We point out that the mean
correlation time is directly accessible in experiments. Note
the logarithmic axis. The triangles denote the melting
concentration of infinitely long random AT and GC stretches,
respectively (from ~\cite{FK}). The maxima of the
$\tau_{\mathrm{corr}}$ curves signify the critical slowing down of the
autocorrelation at the phase transition as before; note that the
maxima coincide with the melting concentrations in the bottom
panel. The dashed line ($\tau_{\mathrm{max}}$ 2D) corresponds to the
longest relaxation time obtained numerically from the ME; it agrees
well with $\tau_{\mathrm{corr}}$ close to the maximum (equivalently
for the other $T$), indicating that at melting there is a single
(slow) relevant relaxation mode.  The horizontal line
($\tau_{\mathrm{max}}$ 1D) represents the analytically obtained
longest relaxation time $(2M+1)^2/ \pi^2 k^{-1}$ for a homopolymer
model, compare Ref.~\cite{jcp}.

The predicted variation with $C$ and $T$ shown in Fig.~\ref{salttemp} is
significant. Thus, different solvent conditions such as temperature or salt
alter the opening probability of the DNA construct, and therefore the blinking
activity. For a fixed salt concentration, for instance, a higher temperature
would therefore lead to more frequent, and longer blinking events, such
that one could measure the effect by both recording individual blinking
events and integrating the blinking signals. As shown in our analysis,
the stability parameters are sufficiently sensitive to externally detect
changes of these parameters. Note that a DNA construct of 30 bps roughly
corresponds to a length of 10nm. Such nanoprobes would easily fit into
nanochannels, small lipid vesicles, or microdishes in gene arrays. We
therefore propose to investigate in more detail the suitability of
DNA-breathing constructs as nanosensors \cite{nanosens,nanosens1}.

\section{Conclusions}

In this study we considered the bubble breathing dynamics in a
heteropolymer DNA-region characterized by statistical weights $u_{\rm
st}(x)$ for disrupting a stacking interaction between neighboring
bps, and the weight $u_{\rm hb}(x)$ for breaking a Watson-Crick
hydrogen bond ($x$ labels different bps), as well the bubble initiation
parameter (the ring-factor) $\xi$. For that purpose, we introduced a
$(2+1)$-variable master equation governing the time evolution of the
probability distribution to find a bubble of size $m$ with left fork
position $x_L$ at time $t$, as well as a complementary Gillespie
scheme. The time averages from the stochastic simulation agree well
with the ensemble properties derived from the master equation. We calculate
the spectrum of relaxation times, and in particular the experimentally
measurable autocorrelation function of a tagged bp is obtained.  All
parameters in our model are known from recent equilibrium measurements
available for arbitrary temperature and NaCl concentration, except for
the rate constant $k$ for (un)zipping that is the only free fit
parameter.
We note that the value for zipping rate obtained from the fluorescence
correlation studies is significantly lower than from NMR experiments
\cite{gueron}. The difference
may stem from the higher temperatures and longer AT sequences probed in the
fluorescence experiments. However, a
perturbing effect of the fluorophore-quencher pair
in the FCS approach cannot be excluded.
For a better understanding of $k$, a more detailed microscopic modeling
and additional experimental study are needed.

We applied recent DNA stability data from \cite{FK,FK1} based on separation of
hydrogen bond and stacking energies. A distinct feature of these parameters
is the low stacking in an TA/AT pair of bps, translating into a pronounced
instability of the TATA motif, as shown for the T7 promoter sequence. We
demonstrated that the probability of simultaneous opening of a stretch of
the size of 4 to 5 bps well discriminates the TATA motif from the other
positions along the promoter sequence, reflecting its biological relevance.
This demonstrates that single DNA fluorescence spectroscopy experiments
can likely be used to investigate in more details the role of the interplay
between TATA-breathing, TATA-box binding proteins, and transcription initiation.
Regarding the mechanism how TATA may guide this initiation we
speculate that it is not primarily the bubble lifetime (much shorter than
the timescale of typical conformational changes of proteins)
but the recurrence frequency of bubble events that
triggers the protein binding.

We note that there exists also a Langevin equation approach to DNA-breathing,
the Dauxois-Peyrard-Bishop model \cite{peyrard,peyrard1}, with seven free
parameters.
Values of these parameters were assigned by comparison to experimental melting
curves for three different short DNA sequences obtained for rather specific
solvent conditions in \cite{campa}. In particular, stacking interactions were
taken to be independent of bp sequence \cite{campa}. In view of the direct
measurement of the stacking free energy in \cite{FK} under various conditions,
it would be desirable to modify the DPB model to accommodate for the full set
of new stability parameters.

We expect this study to encourage furthergoing investigations on the
theoretical understanding of DNA-breathing and the
experimental possibilities to obtain detailed sequence and stability
information of DNA and its interactions with binding proteins from
DNA-breathing dynamics.
We furthermore point out the possibility to
use the results of this study for designing a small
fluorophore/quencher-dressed DNA
construct
for nanosensing applications in nanochannels, vesicles or microdishes.

\begin{acknowledgments}

We thank G.~Altan-Bonnet and A.~Libchaber for sharing the data for
Fig.~\ref{autocorr}, M.~Frank-Kamenetskii for discussion and access to
the new stability data prior to publication, as well as M. A. Lomholt and
K.~Splitorff for
discussion. SKB acknowledges support from Virginia Tech through the ASPIRES
award program.
TA acknowledges partial funding from the Wallenberg foundation.
RM acknowledges partial funding from the Natural Sciences and Engineering
Research Council (NSERC) of Canada and the Canada Research Chairs program.

\end{acknowledgments}

\end{document}